# Generalized Earthquake Frequency-Magnitude Distribution Described by Asymmetric Laplace Mixture Modelling

A. Mignan

*Version date: 1 October 2018*

ETH Zurich, Sonneggstrasse 5, CH-8092 Zurich

Contact: arnaud.mignan@sed.ethz.ch

*Abstract:* The complete part of the earthquake frequency-magnitude distribution (FMD), above completeness magnitude $m_c$, is well described by the Gutenberg-Richter law. The parameter $m_c$ however varies in space due to the seismic network configuration, yielding a convoluted FMD shape below max($m_c$). This paper investigates the shape of the generalized FMD (GFMD), which may be described as a mixture of elemental FMDs (eFMDs) defined as asymmetric Laplace distributions of mode $m_c$ [*Mignan, 2012, https://doi.org/10.1029/2012JB009347*]. An asymmetric Laplace mixture model (GFMD-ALMM) is thus proposed with its parameters (detection parameter κ, Gutenberg-Richter β-value, $m_c$ distribution, as well as *number K* and weight *w* of eFMD components) estimated using a semi-supervised hard expectation maximization approach including BIC penalties for model complexity. The performance of the proposed method is analysed, with encouraging results obtained: κ, β, and the $m_c$ distribution range are retrieved for different GFMD shapes in simulations, as well as in regional catalogues (southern and northern California, Nevada, Taiwan, France), in a global catalogue, and in an aftershock sequence (Christchurch, New Zealand). We find max($m_c$) to be conservative compared to other methods, $k = κ/\log(10) ≈ 3$ in most catalogues (compared to $b = β/\log(10) ≈ 1$), but also that biases in κ and β may occur when rounding errors are present below completeness. The GFMD-ALMM, by modelling different FMD shapes in an autonomous manner, opens the door to new statistical analyses in the realm of incomplete seismicity data, which could in theory improve earthquake forecasting by considering c. ten times more events.

**1 Introduction**

The earthquake frequency-magnitude distribution (FMD) can be represented by the function

$$f(m) = f_{GR}(m)q(m), \tag{1}$$

the product of the Gutenberg-Richter function $f_{GR} = \exp(-\beta m)$ (Gutenberg & Richter, 1944) and a detection function $q(m)$ (e.g. Ringdal, 1975; Ogata & Katsura, 1993; 2006; Mignan, 2012; Alamilla et al., 2014; Kijko & Smit, 2017). The completeness magnitude $m_c$ then represents the magnitude bin at which $q$ tends to 1 and above which the Gutenberg-Richter law prevails. A number of techniques exist to evaluate $m_c$ without requiring knowledge of $q(m)$ (see review by Mignan & Woessner, 2012).

Discarding the seismicity below $m_c$ remains a compulsory step before many seismicity statistical analyses since incomplete data is until now considered unreliable, for lack of understanding of the detection process. Indeed, there is so far no $q$ function that can systematically model the variability of the incomplete part of earthquake catalogues. This is unfortunate, as microseismicity seems to often be required to observe statistically meaningful precursors to large earthquakes (see meta-analysis by Mignan, 2014). At present, two options are possible to increase the amount of available seismicity data, a densification of the seismic network, which represents a long-term endeavour (e.g. Kraft et al., 2013), or seismic waveform template matching, which remains somewhat computationally cumbersome and requires the waveform data (Gibbons & Ringdal, 2006; Shelly et al., 2016). But could we make direct use of the incomplete part of earthquake catalogues, which can represent as much as 90% of the data? We aim at answering this question by offering a $q$ function that is flexible enough that it can model a variety of FMD shapes. This requires defining $m_c$ as a variable instead of a fixed value.

Mignan (2012) presented an earthquake FMD ontology where the FMD is classified by its shape in the log-lin space: Class I '*Angular FMD*', class II/III '*Intermediary FMD with multiple maxima*', class IV '*Gradually curved FMD*', and class V '*Gradually curved FMD with multiple maxima*' (Fig. 1). Class IV is the most commonly studied (e.g. Ringdal, 1975; Ogata & Katsura, 1993; 2006; Kijko & Smit, 2017), as it represents the behaviour of the detected seismicity in relatively large regions in respect to the seismic network spatial extent, such as in regional earthquake catalogues. The gradual curvature has been explained by different $q$ formulations, including the cumulative Normal distribution (Ringdal, 1975; Ogata & Katsura, 1993; 2006) and the generalized gamma distribution (Kijko & Smit, 2017). In contrast, Mignan (2012) showed, in both synthetic and real catalogues, that the gradually curved FMD (class IV) could be represented by the sum of angular FMDs with $q$ an exponential function:

$$q(m; m_c, \kappa) = \begin{cases} \exp[\kappa(m - m_c)] &, m < m_c \\ 1 &, m \geq m_c \end{cases} \quad (2)$$

where $m_c$ is the completeness magnitude and $\kappa$ is a detection parameter. Alamilla et al. (2014; 2015a; 2015b) also proposed, independently, such an exponential function and compared it to the Ringdal normal distribution. Roberts et al. (2015) made the distinction between two types of FMD shapes, 'sharp-peaked' and 'broad-peaked', in agreement with the Mignan ontology.

The so-called 'elemental' angular FMD (class I, or eFMD) is only observed in local earthquake datasets where the detection level is homogeneous, i.e., where $m_c$ is constant (Mignan, 2012; Mignan & Chen, 2016). It is described by the following asymmetric Laplace (AL) probability density function (PDF):

$$p_{AL}(m; m_c, \kappa, \beta) = \frac{1}{\frac{1}{\kappa - \beta} + \frac{1}{\beta}} \begin{cases} \exp[(\kappa - \beta)(m - m_c)] &, m < m_c \\ \exp[-\beta(m - m_c)] &, m \geq m_c \end{cases} \quad (3)$$

where $m_c$ is the mode of the PDF (for the general properties of the AL distribution, see e.g. Kotz et al., 2001). The eFMD was spotted in southern California and Nevada (Mignan, 2012), Greece (Mignan & Chouliaras, 2014), and Taiwan (Mignan & Chen, 2016). The detection functions $q$ defined to fit class IV (e.g. Ogata & Katsura, 2006; Kijko & Smit, 2017) cannot explain the angular shape while a sum of eFMDs can explain an FMD of class IV or V, the shape of gradually curved FMDs being driven by the underlying $m_c$ distribution (Fig. 1).

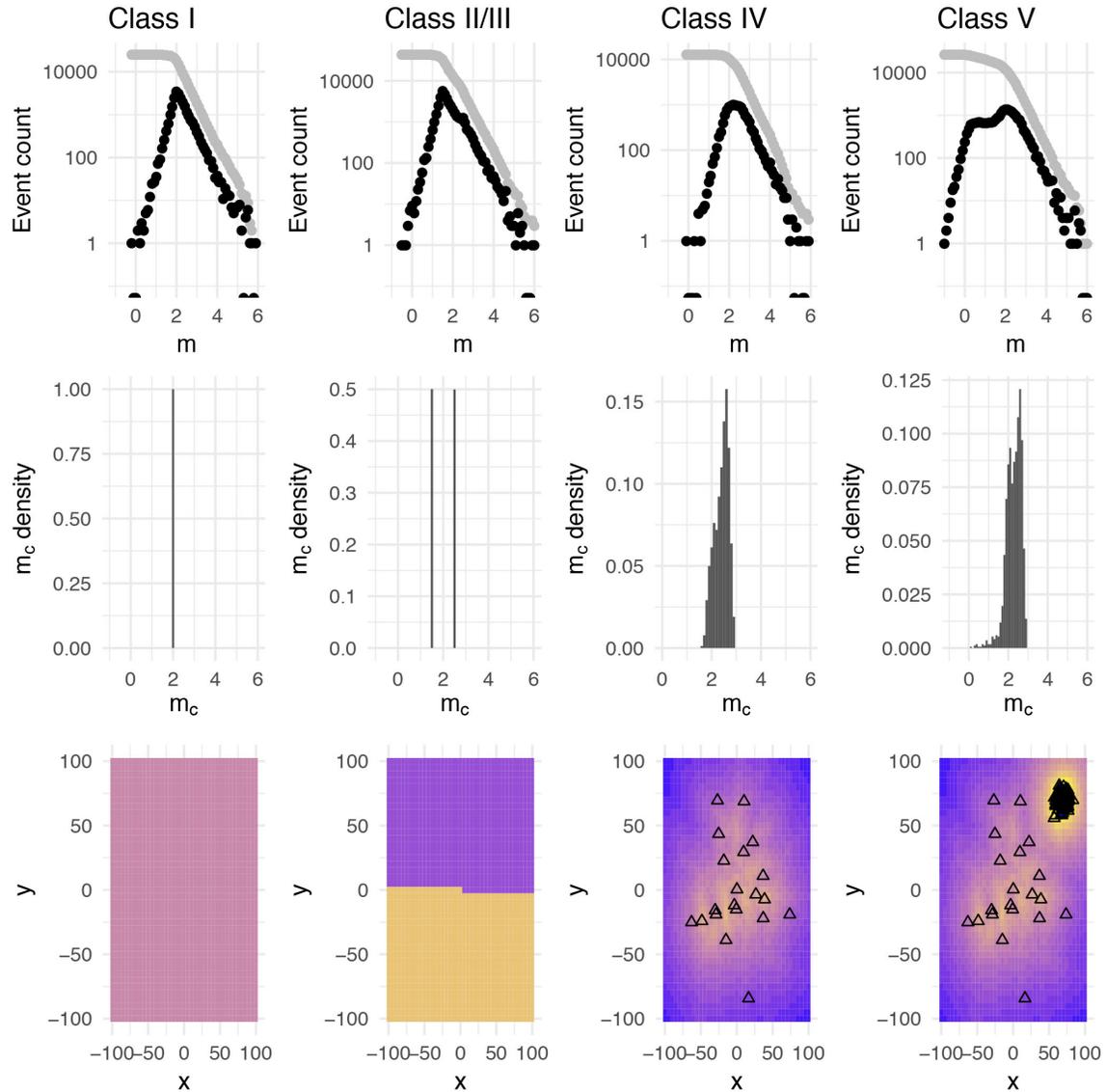

**Figure 1.** Earthquake frequency-magnitude distribution (FMD) ontology, following Mignan (2012). FMDs of different shapes (or classes) are shown in the first row, the matching $m_c$ distributions and maps in the second and third. The FMD classification is as follows: Class I. '*Angular FMD*' due to a constant $m_c$ in space; Class II/III. '*Intermediary FMD with multiple maxima*' due to an increased number of $m_c$ values; Class IV.

'*Gradually curved FMD*' due to the $m_c$ distribution observed at the spatial scale of one seismic network; Class V. '*Gradually curved FMD with multiple maxima*' due to the combination of $m_c$ distributions at different seismic network spatial scales, such as the combination of regional and local networks. The different FMD classes can be represented by the sum of angular FMDs of different $m_c$, with $m_c$ heterogeneities increasing with the class. Low $m_c$ values are represented in yellow and high ones in purple. Seismic stations are represented by open triangles.

The present study offers a generalization of the work initiated by Mignan (2012) by presenting a univariate asymmetric Laplace mixture model (ALMM) that fits the generalized earthquake frequency-magnitude distribution (GFMD), which represents various FMD shapes from class I to V. We will first describe a semi-supervised hard Expectation-Maximization approach that allows fitting the parameters of the GFMD, referred to as GFMD-ALMM (section 2). We will then apply the proposed method on both synthetic catalogues and real data (sections 3-4). We will finally discuss potential applications in earthquake forecasting and possible improvements to the GFMD-ALMM in some concluding remarks (section 5).

## 2 GFMD asymmetric Laplace mixture model (GFMD-ALMM)

*2.1 GFMD simulations*

We first reproduce the earthquake FMD ontology of Mignan (2012) by simulating earthquake catalogues with different $m_c$ distributions (Fig. 1). We assume a number of occurring earthquakes $N_0(m_0 = 0) = 10^3$ per cell (-100 ≤ $x$ ≤ 100 km, -100 ≤ $y$ ≤ 100 km), $\kappa = 3\ln(10)$ and $\beta = \ln(10)$ constant in space (subject to random fluctuation in real catalogues – Mignan, 2012; Kamer & Hiemer, 2015), and $m_c = 2$ for class I, $m_c = \{2, 3\}$ for class III, and $m_c = f_{BMC}(d_i)$ for classes IV and V. In those two last cases,

$$m_c(x,y) = f_{BMC}(d_i) = c_1 d_i(x,y)^{c_3} + c_3 \pm \sigma \qquad (4)$$

with $d_i$ [km] the distance to the $i^{th}$ nearest seismic station, $c_1$, $c_2$ and $c_3$ empirical parameters and $\sigma$ the standard deviation (Mignan et al., 2011 – here using the generic parameters $i = 4$, $c_1 = 5.96$, $c_2 = 0.0803$, $c_3 = -5.80$ and $\sigma = 0$). Eq. (4) is the prior model of the Bayesian Completeness Magnitude (BMC) method and has been validated on the following earthquake catalogues: Taiwan (Mignan et al., 2011), Mainland China (Mignan et al., 2013), Switzerland (Kraft et al., 2013), Lesser Antilles arc (Vorobieva et al., 2013), California (Tormann et al., 2014), Greece (Mignan & Chouliaras, 2014) and Iceland (Panzera et al., 2017). Seismic networks are modelled using a normal distribution, one regional network for class IV ($\mu_x = \mu_y = 0$ km; $\sigma_x = \sigma_y = 30$ km; 20 stations) and two networks with different standard deviations for class V (to mimic the association of a regional network, as above, and a local network: $\mu_x = \mu_y = 70$ km; $\sigma_x = \sigma_y = 5$ km; 100 stations). Note that Eq. (4) represents the first-order $m_c$ fluctuations that one can expect in an earthquake dataset. Temporal changes in the seismic network can be modelled by updating $d_i$. Changes at time $t$ following large earthquakes of magnitude $m$ can be modelled by $m_c(m, t) = m - 4.5 - 0.75\log_{10}(t)$ (Helmstetter et al., 2006). Those second-order temporal changes are not analysed in the present article, although the proposed machine learning approach is agnostic regarding the origin of $m_c$ variations.

Random earthquake magnitudes $\{m_1, \ldots, m_i, \ldots, m_N\}$ are simulated from the angular FMD function for each cell $(x, y)$ by applying the Inversion Method (Devroye, 1986; Clauset et al., 2009) to the complementary cumulative density function, obtained by integrating Eq. (3)

$$P_{AL}(m) = \int_m^\infty p_{AL}(m')dm' = 1 - u = \begin{cases} \exp[(\kappa - \beta)(m - m_c)] &, m < m_c \\ \exp[-\beta(m - m_c)] &, m \geq m_c \end{cases} \quad (5)$$

where $u$ is a random number uniformly distributed in the interval (0, 1). It yields the random variable

$$m_i = \begin{cases} m_c + \frac{1}{\kappa - \beta}\ln(1 - u) &, m < m_c \\ m_c - \frac{1}{\beta}\ln(1 - u) &, m \geq m_c \end{cases} \quad (6)$$

Due to the piecewise nature of Eq. (3), the total number of events to simulate per cell, $N = N(m \geq m_c) + N(m < m_c)$, is

$$\begin{cases} N(m \geq m_c) = N_0(m_0)\exp(-\beta[m_c - m_0]) \\ \quad N(m < m_c) = \frac{\beta}{\kappa-\beta} N(m \geq m_c) \end{cases} \quad (7)$$

The GFMD is finally the sum of all angular FMDs over all cells ($x$, $y$). Figure 1 shows the results for the different tested $m_c$ distributions, with the different classes of the Mignan (2012) FMD ontology retrieved: classes I ("*angular*"), II/III ("*intermediary with multiple maxima*"), IV ("*gradually curved*") and V ("*gradually curved with multiple maxima*"). Class II, a special case of class III in which two maxima are separated by one magnitude bin Δ$m$, is not investigated.

*2.2 ALMM fitting by semi-supervised hard Expectation-Maximization*

The asymmetric Laplace mixture model (ALMM) of the earthquake GFMD is defined by the following probability density function

$$p_{ALMM}(m; m_{ck}, \kappa_k, \beta_k) = \sum_{k=1}^{K} w_k p_{AL}(m; m_{ck}, \kappa_k, \beta_k) \quad (8)$$

with $K$ the number of angular FMDs, $m_{ck}$, $\kappa_k$ and $\beta_k$ the parameters of the $k^{th}$ FMD component, and $w_k$ the mixing weight of the $k^{th}$ component such that $\sum_{k=1}^{K} w_k = 1$. The ALMM is flexible as it is able to fit diverse FMD patterns from class I to V. For illustration purposes, Figure 2 represents both the AL eFMD mixture components (in orange) and the ALMM's GFMD (in red) for a simulated FMD of class IV. The $m_c$ distribution is estimated from an $m_c$ spatial map (Fig. 1; Eq. 4), and the mixing weight $w_k(m_{ck})$ (in brown) corresponds to the sum of $N_k$ over cells of completeness magnitude $m_{ck}$, normalized by $N_{tot}$, the total number of events. Note that κ and β are considered constants in the ALMM for the rest of this paper (in agreement, at first-order, with Mignan, 2012; Kamer & Hiemer, 2015). Magnitudes are binned in Δ$m$ = 0.1 intervals.

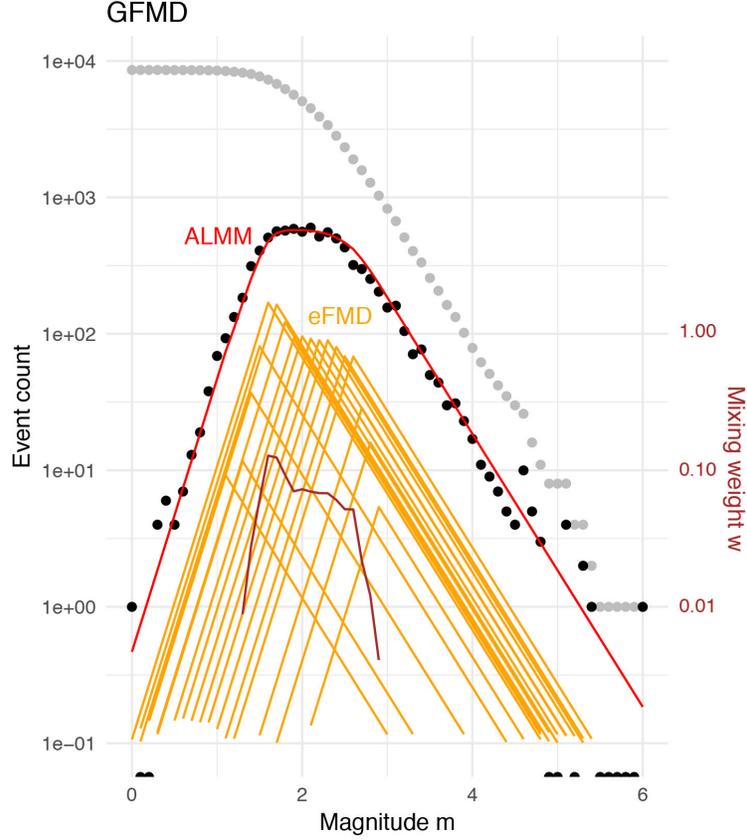

**Figure 2.** Generalized earthquake frequency magnitude distribution (GFMD) of class IV, defined as the sum of elemental angular FMDs (eFMDs; Eq. 3; in orange) and described by the asymmetric Laplace mixture model (ALMM; Eq. 8; in red). The distribution of the mixing weights $w(m_c)$ controls the shape of the GFMD, as demonstrated in Figure 1.

The ALMM can be fitted to different FMD shapes by using Expectation-Maximization (EM), a machine learning class of iterative algorithm for clustering (Dempster et al., 1977; Redner & Walker, 1984; Moon, 1996). Many variants of the EM algorithm exist (e.g. Samdani et al., 2012). The approach presented below is a simple case of hard EM, as our goal is not to assign the probability of having $m$ belonging to a given cluster, but to define a surrogate of the true (but unknown) GFMD in order to estimate the value of $K$, $w_k$, $m_{ck}$, $\kappa$ and $\beta$. The proposed EM algorithm, applied for $K = \{1, 2, …, K_{max}\}$ components, is defined as follows:

We set the initial parameter values $m_{ck}$, $\kappa$ and $\beta$ by applying k-means (MacQueen, 1967; Jain, 2010), with $k = \{1, 2, …, K\}$, $w_k$ the normalized number of events per cluster, and $m_{ck}$ the cluster centre. The magnitude vector is defined as $\mathbf{M} = \{\mathbf{M}_1, \mathbf{M}_2, …, \mathbf{M}_K\}$, ordered by increasing $m_{ck}$ and with each component defined as $\mathbf{M}_k = \{m_1, m_2, …\}$, the feature vector of magnitude scalars $m$ to be labelled to cluster $k$. We obtain parameters κ and β from the clusters of centres $m_{c1} = \min(m_{ck})$ and $m_{cK} = \max(m_{ck})$, or $\mathbf{M}_1$ and $\mathbf{M}_K$, respectively, by using the maximum likelihood estimation (MLE) method:

$$\begin{cases} \chi = 1/\left(\left(\min(m_{ck}) - \frac{\Delta m}{2}\right) - \overline{\mathbf{M}}_{left}\right) \\ \beta = 1/\left(\overline{\mathbf{M}}_{right} - \left(\max(m_{ck}) - \frac{\Delta m}{2}\right)\right) \end{cases} \quad (9)$$

where $\chi = \kappa - \beta$ is the slope of the incomplete part of the eFMD, $\mathbf{M}_{left} = \{m \in \mathbf{M}_1 : m \leq m_{c1} - \frac{\Delta m}{2}\}$ and $\mathbf{M}_{right} = \{m \in \mathbf{M}_K : m > m_{cK} - \frac{\Delta m}{2}\}$ (Aki, 1965). Although k-means may provide biased estimates of $m_{ck}$, it nevertheless reliably finds the local maxima of the $m$-space (hence avoiding one of the common difficulties of mixture modelling; e.g. Celeux et al., 2000). We do not use k-modes, as it is defined for categorical data (Huang, 1997). We will however refine the $m_{ck}$ values using the mode in an iterative EM.

At each iteration $i$, a deterministic version of the expectation step (E-step) attributes a hard label $k$ to each $m$-event from the parameter set $\theta_k^{[i-1]} = \{m_{ck}, \kappa, \beta\}$ defined in the previous iteration $i$-1 ($i = 0$ corresponding to the k-means estimates). Hard labels are assigned as:

$$k = \operatorname{argmax}_k p_{AL}(\theta_k^{[i-1]}, m) \quad (10)$$

The maximization step (M-step) updates the component parameters: $w_k$ is the normalized number of $m$-events per component $k$, $m_{ck} = \operatorname{mode}(\mathbf{M}_k)$, and κ and β are calculated from Eq. (9). It should be mentioned that when $m_{ck}$ extrema lead to component under-sampling and therefore to errors in κ and β, $\mathbf{M}_{1+j}$ and $\mathbf{M}_{K-l}$ are used instead in Eq. (9) (with $j$ and $l$ increased incrementally until no error is found). For classes I to III GFMDs, $m_{ck}$ estimates rapidly fall into the local FMD maxima. However, for class IV GFMDs, the estimates tend to migrate towards the unique maximum. This problem is avoided by shifting $m_{ck}$ to the nearest free magnitude bin when a bin is already occupied for the class IV/V case (some sort of semi-supervised clustering; e.g. Jain, 2010). To determine whether a GFMD is part of class IV/V or not, we first apply a classifier 'curved/not-curved' at iteration $i = 1$ (see below).

The E- and M-steps are repeated until log-likelihood $LL$ convergence (difference between two iterations lower than $10^{-6}$) or until $i = i_{max} = 5$ (a higher $i_{max}$ does not significantly improve the results). Once the procedure has been repeated $K_{max}$ times, the best number of components is $K_{BIC}$, the number of components with the lowest Bayesian Information Criterion (BIC) estimate BIC($K$) = $-LL+1/2 \cdot n_{par} \ln N_{tot}$ with $n_{par} = 2+K$ (Schwarz, 1978; with 2 representing the free parameters κ and β). Note that computing the log-likelihood from the function $\sum \ln(p_{ALMM}(m))$ is inconclusive due to higher weights on $m_{ck}$ components with the largest $\mathbf{M}_k$ size. To avoid this bias towards the main mode of the distribution, we compute instead the log-likelihood of a Poisson process:

$$LL(\theta_k^{[i]}, X = \{n_j; j = 1, \ldots, N_j\}) = \sum_{j=1}^{N_j}\left[n_j \ln(v_j(m_j|\theta_k^{[i]})) - v_j(m_j|\theta_k^{[i]}) - \ln(n_j!)\right]$$

(11)

for the observed magnitude rate $n_j \left(m \in \left(m_j - \frac{\Delta m}{2}, m_j + \frac{\Delta m}{2}\right]; m_j = 0.0, 0.1, \ldots, 8.0\right)$ and predicted rate

$$v(m|\theta_k^{[i]}) = N_{tot} p_{ALMM}(m; \theta_k^{[i]}) \Delta m \qquad (12)$$

Hence the present MLE method is an estimator of the shape of the GFMD represented by the rate ν($m$), instead of the population of magnitudes $m$. This assumes that the temporal clustering of earthquake has no effect on the FMD model, which remains questionable since large earthquakes alter the completeness $m_c$ shortly after their occurrence (Helmstetter et al., 2006; Mignan & Woessner, 2012). As for the 'curved/not-curved' classifier, we compare the BIC estimates of the mixture model (Eq. 12) and of the curved FMD model of Ogata & Katsura (2006) (see their Eq. 6) using the $LL$ definition of Eq. (11). If the FMD is not curved, the mixture model will lead to a lower BIC; however, if the FMD is curved, the mixture model will lead to a higher BIC since, even if it would fit reasonably well the GFMD shape at $i = 1$, it is penalized for the higher number of parameters compared to the simple 3-parameter Ogata-Katsura model.

Finally note that in the cases where the EM algorithm fails for a given stochastic realisation, this realisation is not recorded (the k-means iteration would always provide a result, but likely biased). The EM algorithm may fail in convoluted cases (e.g. class V simulations) and more so in real cases. However, we did not come upon an FMD shape in which the GFMD-ALMM would systematically fail (see results below).

**3 Data**

The GFMD ALMM is first tested on simulations, as described in section 2.1. In the case of real data, in order to compare the mixture model results with previous studies, we first use the southern California and Nevada earthquake catalogues as defined in Mignan (2012), which is the only study available with estimates of the detection parameter κ. Mignan (2012) estimated this parameter (as well as β and $m_c$) from the fitting of the eFMD (Eq. 3) in grids of relatively high resolution to minimize $m_c$ heterogeneities (and hence the FMD curvature). For southern California, the data originates from the Southern California Earthquake Data Center (SCEDC, 2013) for the period 2001-2007 (inclusive). For Nevada, the catalogue had been retrieved from the Advanced National Seismic System (ANSS) composite catalogue (NCED*C*, 2016) for the period 2000-2009 (inclusive). Both catalogues were constrained in space by their respective ANSS authoritative regions (available at http://www.ncedc.org/anss/anss-detail.html#regions, last assessed May 2018) and a maximum 20-km depth. The Nevada catalogue is of particular interest here as it represents a class V dataset (Fig. 1), resulting from the combination of the regional Nevada Seismic Network and the local Southern Great Basin Digital Seismic Network centred on Yucca Mountain.

Six additional earthquake catalogues, all available in the literature, are considered in order to investigate how the proposed mixture model generalizes. Regional catalogues are from southern California (Hauksson et al., 2012), northern California (Waldhauser & Schaff, 2008), Taiwan (Wu et al., 2008), and France (Cara et al., 2015). We also test the ALMM on the ISC-GEM Global Instrumental Earthquake Catalogue (Storchak et al., 2013) and the 2011 $M_w$6.2 Christchurch, New Zealand, aftershock sequence (Bannister et al., 2011). All datasets are used 'raw', as published.

## 4 Testing of the GFMD-ALMM

*4.1 ALMM simulation results*

We first test the ALMM-GMFD on simulated FMDs of class I/III (as defined in section 2.1). Figure 3 shows four examples of GFMDs where the true $m_{ck}$ values are regularly spaced and with equal weights $w_k$, for $K_{true}$ = 1 to 4. The first column shows one GFMD example taken from 1,000 simulations, the true model (black dotted curves), and the ALMM-GMFD fit (eFMDs in orange and their mixture in red). The central column shows the distribution of $K_{BIC}$ for the 1,000 simulations, and the third column the obtained $\theta_k$ distribution. True values of $K$ and $\theta_k$ are represented by vertical lines. Our GFMD-ALMM fitting procedure retrieves $K$ and $\theta_k$ reasonably well. Note that using random weights only alters $K_{BIC}$, which decreases when the EM algorithm does not find components with very low weight $w_k$. This becomes systematic in more realistic $m_c$ distributions, as shown below.

We then test the proposed method on FMDs of class IV/V (class IV represents the most common case, the so-called 'bulk FMD' or 'regional FMD', whereas class V remains relatively rare, when both regional and local networks are present). The $m$ vector is simulated using the method described in section 2.1 where the true $m_{ck}$ distribution depends on the seismic network spatial configuration (Eq. 4). Examples of simulated $m_c$ maps for class IV and class V are shown in the top row of Figure 4. One example of GFMD out of 1,000 simulations is shown on the second row for those two classes. The colour coding is the same as in Figure 3. The GFMD-ALMM retrieves the curved and curved-with-two-maxima FMD shapes reasonably well. The $K_{BIC}$ and $\theta_k$ distributions are shown on the third and fourth rows with the true values represented by vertical lines. The parameter set $\theta_k$ is again reasonably well recovered although κ is slightly underestimated for class V; note also the bimodal $m_{ck}$ distribution obtained for class V, which is representative of this convolute class. In contrast to previous tests (Fig. 3), the number of components $K$ is now systematically underestimated, which can be explained by the presence of low-weight eFMD components in realistic $m_c$ distributions. This however does not seem to have a significant impact on the estimation of $\theta_k$.

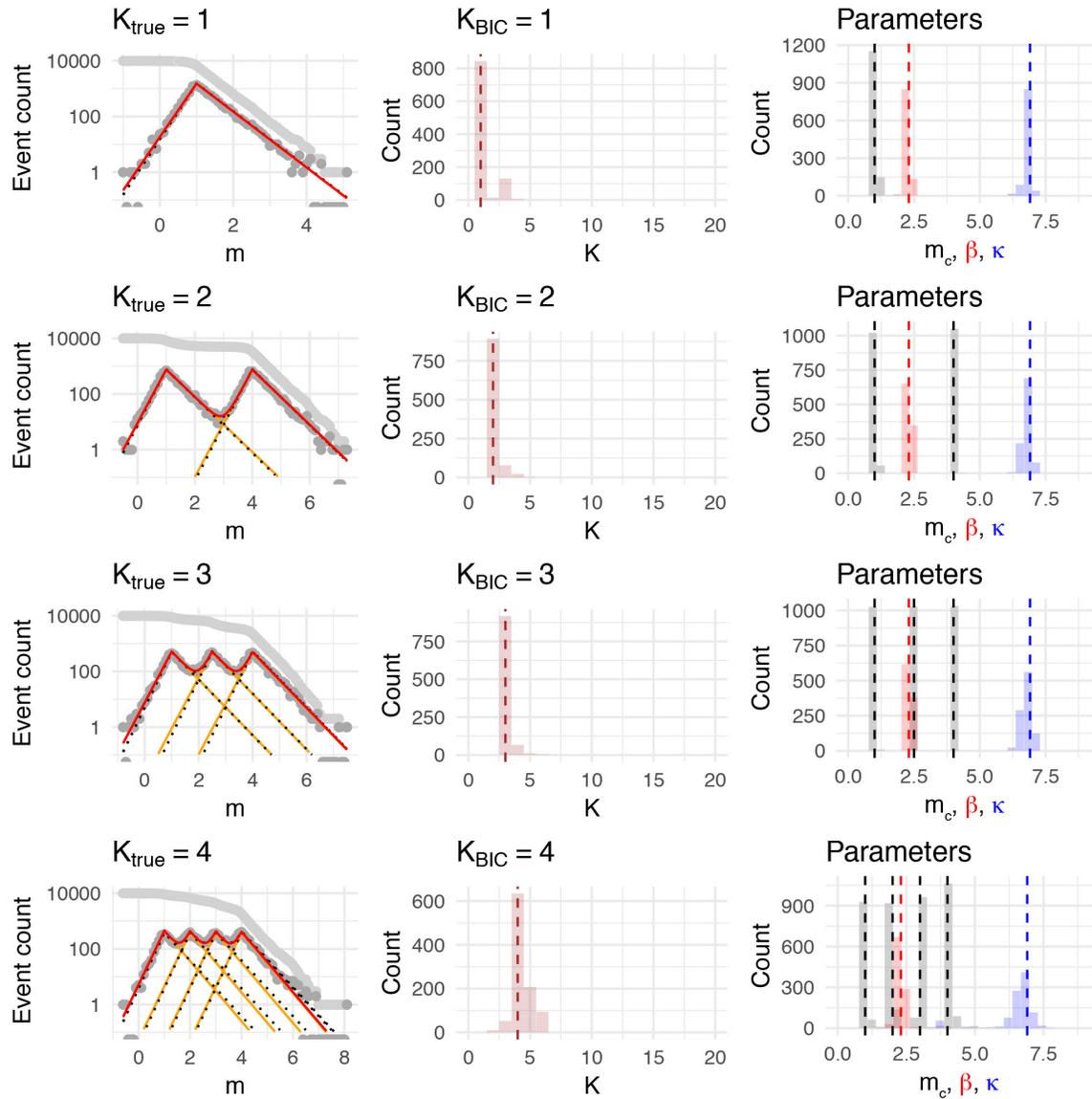

**Figure 3.** ALMM fitting results for class I/III GFMDs. The first column shows one GFMD example from 1,000 simulations, the true model (black dotted curves), and the simulation fit (eFMDs in orange and their mixture in red). The central column shows the distribution of $K_{BIC}$ for the 1,000 simulations, and the third column the $\theta_k$ distribution. True values of $K$ and $\theta_k$ are represented by vertical dashed lines.

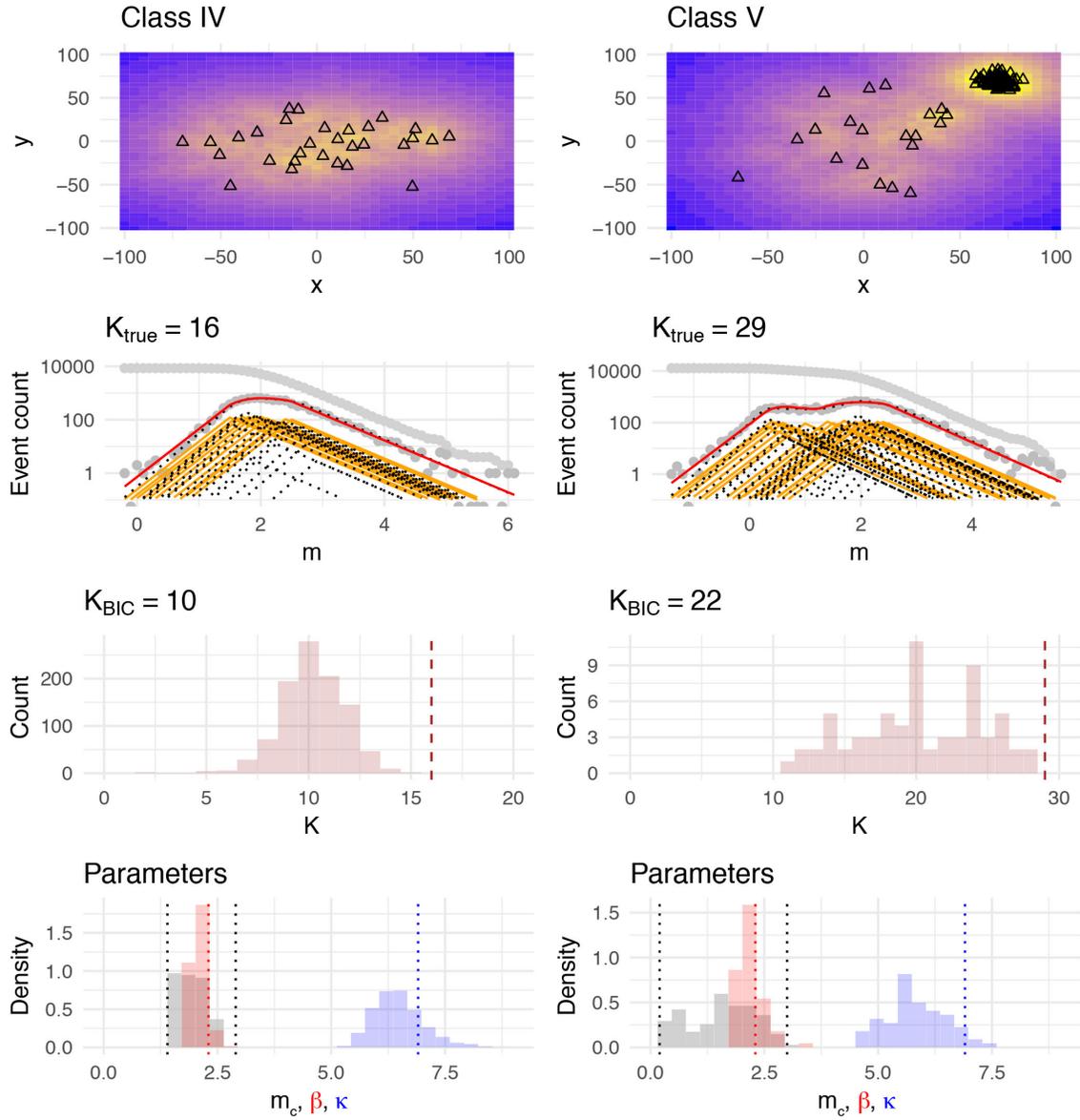

**Figure 4.** ALMM fitting results for simulated class IV and class V GMFDs. The first row shows the $m_c$ maps of the synthetic seismic networks represented by the open triangles (Eq. 4). Low $m_c$ values are represented in yellow and high ones in purple. The second row represents one GFMD example from 1,000 simulations, with the true model (black dotted curves), and the simulation fit (eFMDs in orange and their mixture in red). The third and fourth rows show the distributions of $K_{BIC}$ and $\theta_k$, respectively. True values of $K$ and $\theta_k$ are represented by vertical lines. In this case, $K$ is underestimated but can be retrieved by other means (see Fig. 5).

$K$, which can be seen as a proxy to the degree of $m_c$ heterogeneity, can however be estimated independently if needed. The BMC prior model (Eq. 4) suggests that $m_c$ evolves faster in the dense part of the seismic network than in the sparser areas. Mignan et al. (2011) made use of that observation to optimize the size of the regions where $m_c$ should be homogeneous. Mignan & Chen (2016) proved this trend independently of the BMC prior model showing that the eFMD is only observed in small volumes where the seismic network is dense while it can be observed in larger volumes away from the network. From Eq. (4), we can therefore derive

$$L(K, d_i) = \left(\frac{c_1 d_i^{c_2} + \left(K - \frac{3\sigma}{\Delta m}\right)\frac{\Delta m}{2}}{c_1}\right)^{\frac{1}{c_2}} - \left(\frac{c_1 d_i^{c_2} - \left(K - \frac{3\sigma}{\Delta m}\right)\frac{\Delta m}{2}}{c_1}\right)^{\frac{1}{c_2}} \quad (13)$$

where $d_i$ is the distance to the $i^{th}$ nearest seismic station and $L = 2\sqrt{A/\pi}$ is the characteristic length of the area of interest $A$ (see a simplified expression in Mignan et al. (2011) originally proposed to minimize $m_c$ heterogeneities in BMC mapping).

Figure 5 shows how $K$ depends on $d_5$ (distance to the 5$^{th}$ nearest seismic station) and $L$ for simulations of $m_c$ in a regional network (as defined in section 2.1). As in Mignan & Chen (2016), we estimate $d_5$ from the nucleus of Voronoi cells (Voronoi, 1908; Lee & Schachter, 1980) and $L$ from the Voronoi cell area $A$ (see example of Voronoi tessellation on the left column). The true $K$ is estimated as the sum of unique $m_c$ values per cell, constrained by the bin $\Delta m$. The resulting $K(d_5, L)$ distribution, obtained for 100 Voronoi tessellations, is plotted on the right column of Figure 5 with $\sigma = 0$ in the top row (no noise) and $\sigma = 0.18$ in the bottom row (with noise; value taken from the generic prior BMC model; Mignan et al., 2011). Note that we retrieve the same spatial scaling of detected seismicity as described by Mignan & Chen (2016), but instead of mapping a 'curved versus angular' criterion, we directly map the degree of FMD curvature, and therefore of $m_c$ heterogeneity, from $K$. The curves that represent Eq. (13) for different $K$ values are consistent with the simulation results, hence proving that $K$ can be approximated independently of the GFMD-ALMM.

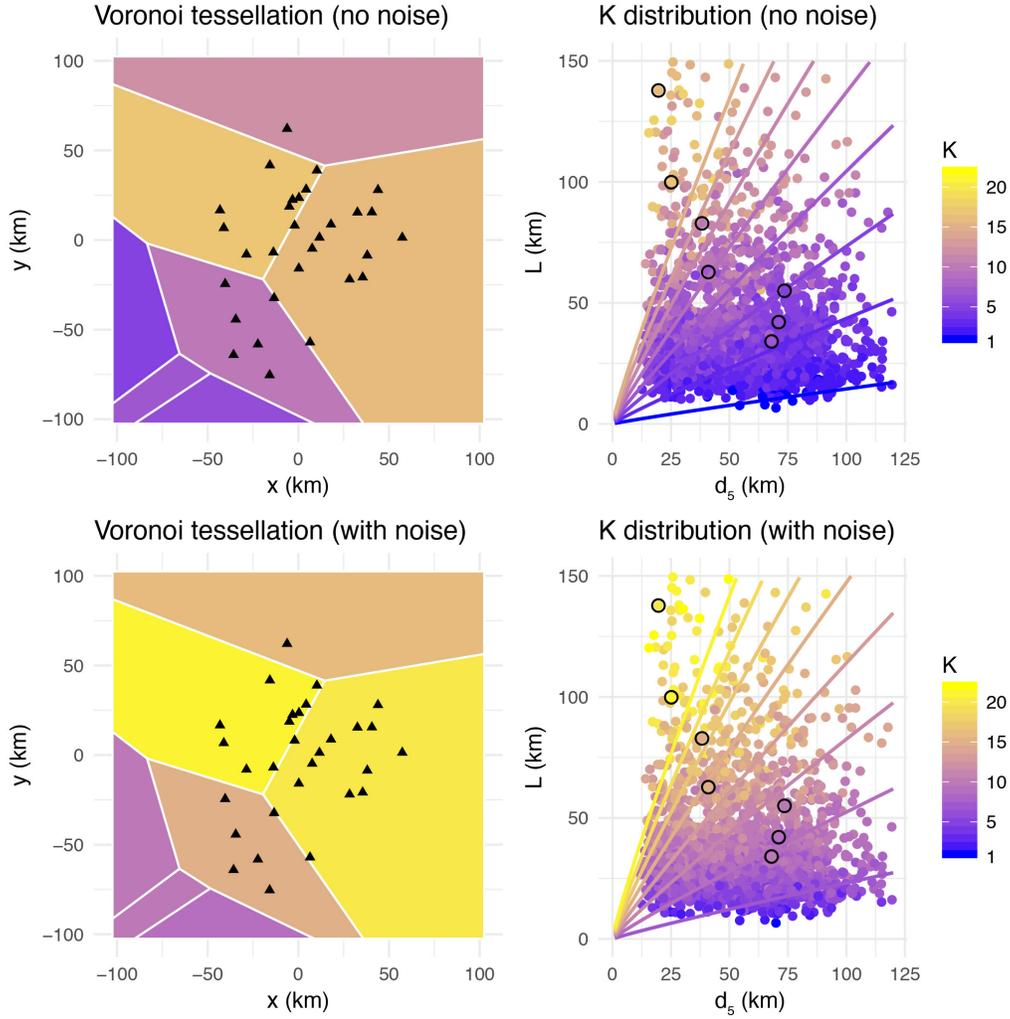

**Figure 5.** Spatial scale of detected seismicity described by the number of mixture components $K$, as a function of the distance to the 5$^{th}$ nearest seismic station $d_5$ and of the Voronoi cell characteristic length $L$. Seismic stations are represented by triangles, in map view (left). For each cell, $d_5$ is calculated from the Voronoi nucleus coordinates and $L$ from the cell area $A$ with $L = 2\sqrt{A/\pi}$. $K$ is estimated for each cell as the number of unique $m_c$ values in that cell, simulated following Eq. (4) with $\sigma = 0$ in the top row (no noise) and with $\sigma = 0.18$ in the bottom row (with noise). The $K(d_5, L)$ distribution of 100 Voronoi models (right) is represented by dots (the circled dots represent the results of the Voronoi model instance shown on the left column). The $K(d_5, L)$ distribution is reasonably well predicted by Eq. (13), represented by curves for different $K$ values.

So far, the GFMD-ALMM does not exploit the BMC prior as side information. One could imagine better constraining the mixture modelling by the predicted $K(d_i, L)$ for a given seismic network and spatial area. The FMD ontology (Mignan, 2012; Fig. 1) could hence be used to improve the semi-supervised clustering with additional 'must-link' and 'cannot-link' constraints specified (e.g. Jain, 2010).

*4.2 Comparison of the ALMM results with the eFMD mapping results of Mignan (2012)*

We now test the applicability of the GFMD-ALMM on the southern California and Nevada catalogues originally investigated in Mignan (2012). Figure 6 shows the results. The first row shows the bulk FMD of southern California and Nevada, respectively (grey dots). The ALMM's GFMD is represented in red with the eFMD components in orange, as in previous figures. We first note that the curved shape of the southern California FMD (class IV) and the curved shape with two maxima of the Nevada FMD (class V) are relatively well approximated by the model. To estimate the accuracy of the GFMD-ALMM as done previously with simulations, we bootstrap the real data 100 times (Efron, 1979; 2003). The resulting distribution of $m_c$ is shown in the second row and the distributions of β and κ in the third row. We get $m_{c,SC}$ = 1.46±0.35 in the min/max range [0.6, 2.3], $κ_{SC}$ = 6.34±0.66 ($k_{SC}$ = 2.76±0.29 in $\log_{10}$ scale) and $β_{SC}$ = 2.52±0.18 ($b_{SC}$ = 1.09±0.08 in $\log_{10}$ scale) for southern California, and $m_{c,NV}$ = 0.47±0.64 within the min/max range [-0.7, 1.7], $κ_{NV}$ = 9.58±1.89 ($k_{NV}$ = 4.16±0.82 in $\log_{10}$ scale) and $β_{NV}$ = 2.09±0.22 ($b_{NV}$ = 0.91±0.09 in $\log_{10}$ scale) for Nevada. Note that the bimodal $m_c$ distribution of the Nevada catalogue is well retrieved, with the two main modes -0.2 and 1.1 representative of the local and regional networks, respectively.

The ranges of mean values obtained by Mignan (2012) for 0.2°, 0.1° and 0.05° spatial resolutions are represented by solid rectangles in Figure 6 and the 1-sigma ranges by open rectangles (see their Table 2). The GFMD-ALMM results are overall compatible with the values provided by Mignan (2012) despite the different approaches: here we estimate the mixture parameters directly from the bulk FMD while Mignan (2012) fitted the eFMDs individually at the local level, assuming $K$ = 1, a reasonable yet approximate approach for the range $5 \leq L \leq 20$ km (Fig. 5). Comparing the lower part of the bulk FMDs below min($m_c$) (top row), one can note that κ is well constrained for Nevada but is slightly underestimated for southern California due to the truncation at *m* = 0. This would explain why $\kappa_{SC}$ is in the low range of estimates obtained by Mignan (2012) in this specific case.

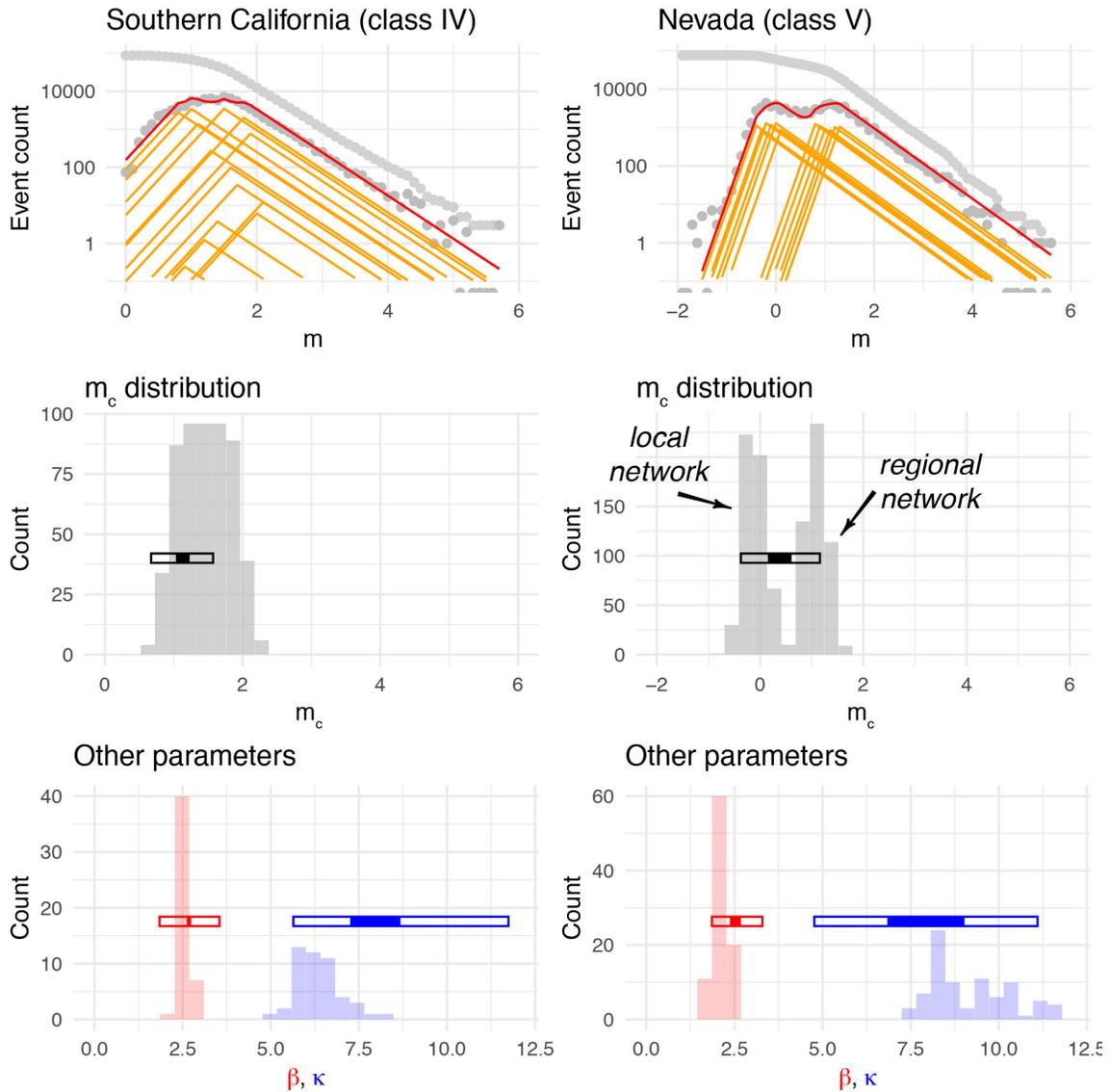

**Figure 6.** Comparison of the ALMM with the results of Mignan (2012). The first row shows the southern California bulk FMD of class IV and the Nevada bulk FMD of class V, with examples of GFMD-ALMM fits (eFMDs in orange and their mixture in red). The second and third rows show the distributions of $m_c$ and of β and κ, respectively, for 100 bootstraps of the data. Those values are compared to the results obtained independently by Mignan (2012) for the same data (their mean ranges and 1-sigma ranges represented

by solid and open rectangles, respectively – different ranges of values had been obtained for different spatial resolutions; see text for details).

*4.3 Comparison of ALMM parameter estimates with results of other FMD models*

We additionally test the GFMD-ALMM on the six published earthquake catalogues listed in section 3 (Bannister et al., 2011; Cara et al., 2015; Hauksson et al., 2012; Storchak et al., 2013; Waldhauser & Schaff, 2008; Wu et al., 2008). Results are shown in Figure 7 for southern California, northern California and Taiwan, and in Figure 8 for France, the ISM-GEM global catalogue and the 2011 Christchurch aftershock sequence. The first row shows the bulk FMD (grey dots), the ALMM's GFMD (in red) and the eFMD components (in orange). The second and third rows show the respective parameter distributions for 100 bootstraps per dataset. Despite differences between the six catalogues ('sharp-peaked' versus 'broad-peaked', sample size ranging from c. 2,000 to 500,000 events, local versus global), the proposed mixture model retrieves the various FMD shapes reasonably well. For the aftershock sequence special case, we do not find the Poisson approximation (Eq. 11) to have an impact on the fitting despite the $m_c$ time-dependency. The parameter estimates of the ALMM are listed in Table 1, with β and κ given in log(10) scale (i.e., *b*- and *k*-values).

We also compare the $m_c$ and *b* estimates obtained by the ALMM to the FMD model of Ogata & Katsura (1993; 2006) and to a non-parametric FMD-based $m_c$ estimator, the median-based analysis of the segment slope or MBASS (Amorèse, 2007). In particular, we compare our bulk estimate $\max(m_{ck})$ to $\mu_{OK}+n\sigma_{OK}$ and $m_{c,MBASS}+n\sigma_{MBASS}$, where $\mu_{OK}$ and $\sigma_{OK}$ are the mean and standard deviation of the cumulative Normal distribution of the Ogata-Katsura detection function, and $m_{c,MBASS}$ and $\sigma_{MBASS}$ are the mean and standard deviation of the MBASS estimates for 100 bootstraps. A high *n* represents a more conservative estimate of $m_c$. $b_{MBASS}$ is estimated from the Aki (1965) method for MBASS $m_c$ values whereas $b_{OK}$ is estimated by the Ogata-Katsura model. Results are shown in Figure 9, in purple for the Ogata-Katsura model, in dark green for MBASS, and with decreasing dash length for increasing *n*. Parameter estimates are given in Table 1 for conservative estimates only (case *n* = 3).

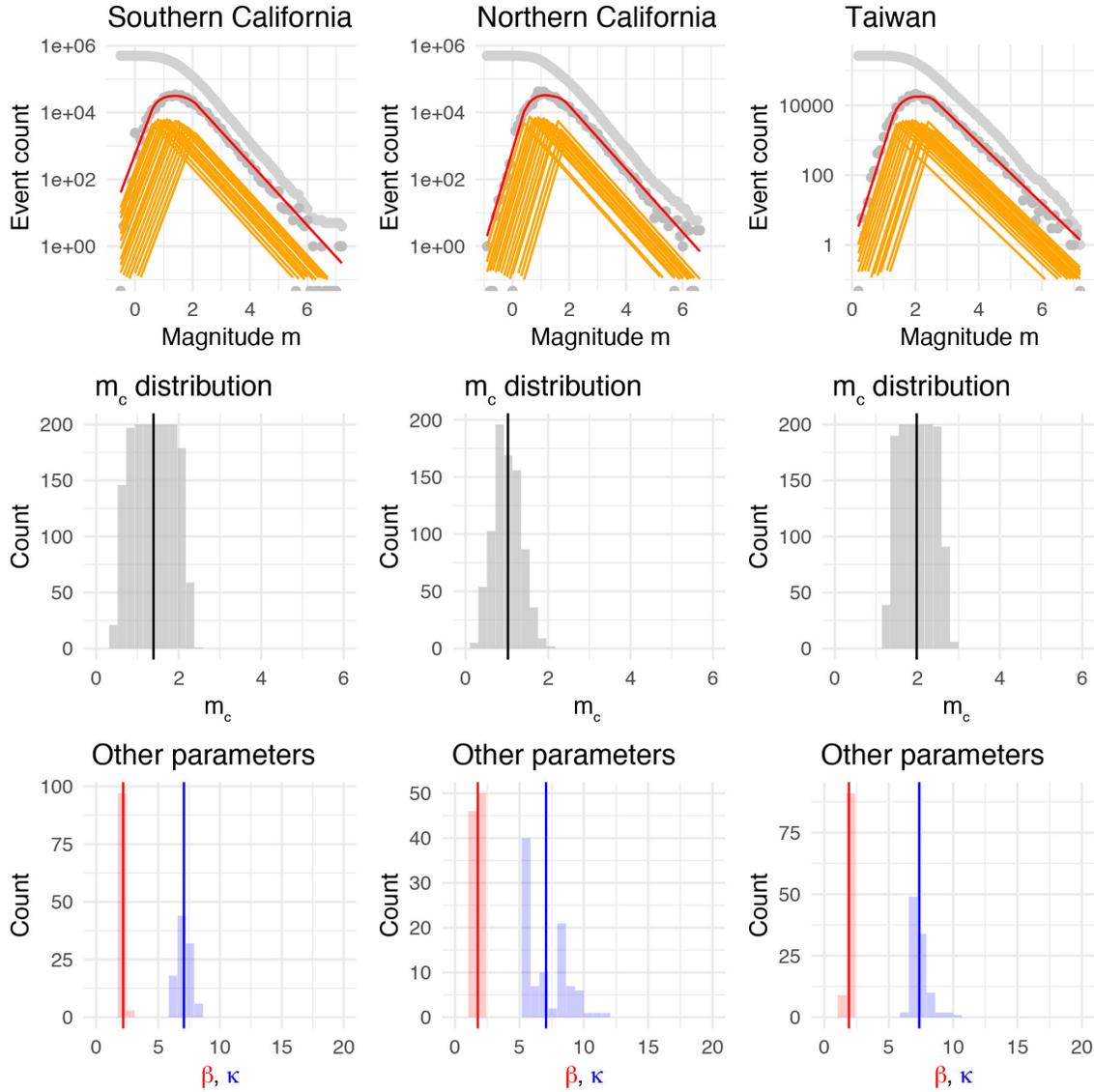

**Figure 7.** Application of the ALMM to southern California, northern California and Taiwan. The first row shows the observed bulk FMDs, with examples of GFMD-ALMM fits (eFMDs in orange and their mixture in red). The second and third rows show the distributions of $m_c$ and of β and κ, respectively, for 100 bootstraps. The vertical solid lines represent the mean parameter estimates. Note that events of magnitude $m = 0.0$, due to their abnormally high number, were here removed from the northern California catalogue (see text for details, as well as Fig. 9 for the original $m=0$ peak).

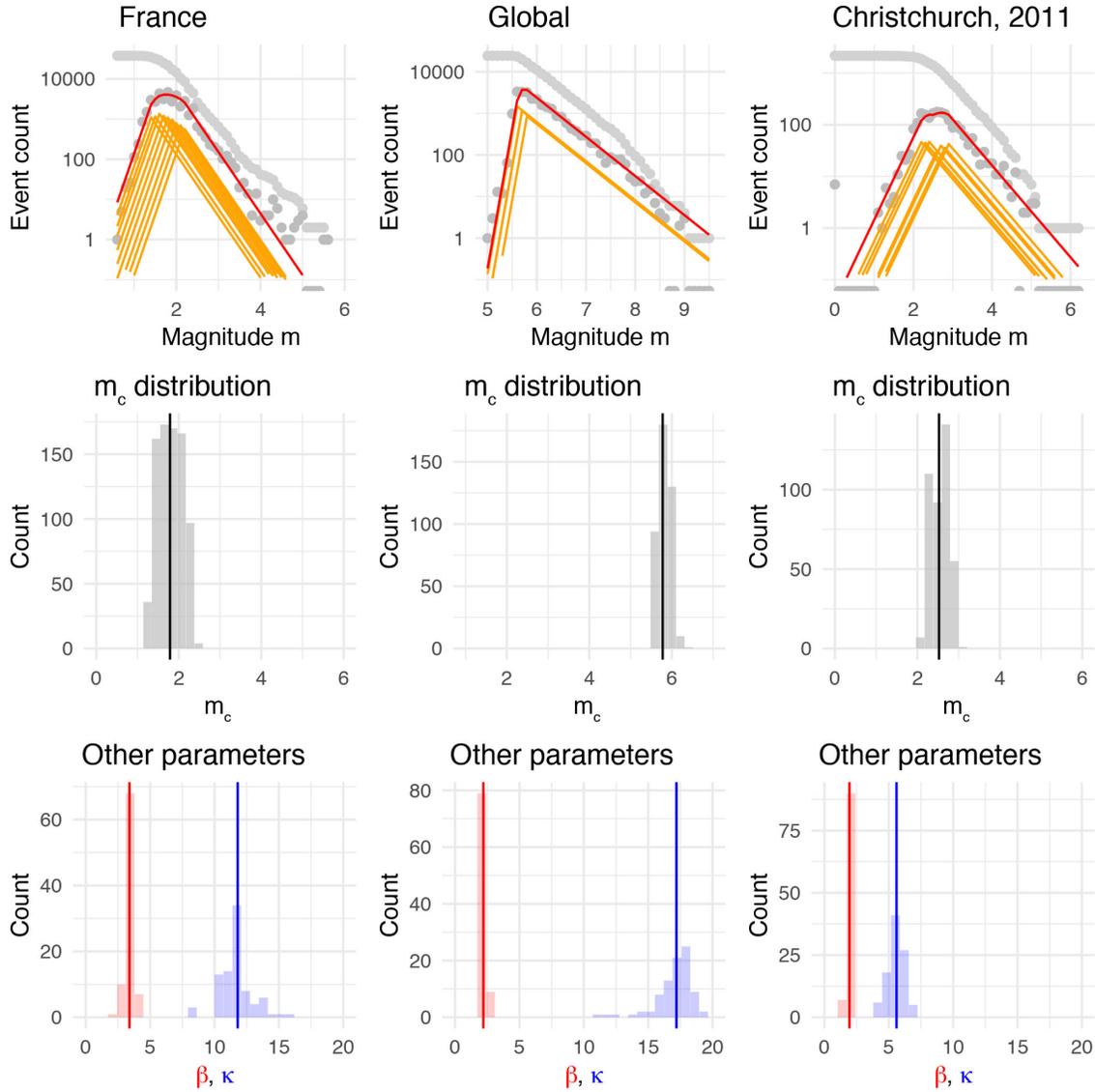

**Figure 8.** Application of the ALMM to France, the ISC-GEM global catalogue and the 2011 Christchurch aftershock sequence. The first row shows the observed bulk FMDs, with examples of GFMD-ALMM fits (eFMDs in orange and their mixture in red). The second and third rows show the distributions of $m_c$ and of $\beta$ and $\kappa$, respectively, for 100 bootstraps. The vertical solid lines represent the mean parameter estimates.

We first note a general agreement between methods, except for northern California and France where the ALMM mean *b*-value is too low and too high, respectively. This demonstrates the possible impact of the *k*-value (=$\kappa$/log(10)) on *b* since it is in those two cases that *k* uncertainty is the highest with a standard deviation of c.

0.70 instead of c. 0.25. This high uncertainty, and in the case of northern California the bimodal distribution of $k$ (Fig. 7), may be due to apparent problems with the incomplete part of the FMD. For northern California, the ALMM fails on the original catalogue due to an anomalous peak at $m = 0.0$, visible in Figure 9. Removing all $m = 0$ events lets the ALMM work but at the expense of higher uncertainty on the FMD shape. Similarly, there seems to be rounding errors in the SI-Hex catalogue of France, represented by a 'zig-zag' pattern at the top of the FMD. This renders the ALMM fitting unstable. Therefore, the ALMM seems sensitive to possible rounding errors below completeness. Those might be neglected by seismic network operators since the data below $m_c$ is usually discarded.

We then find that the ALMM bulk estimate max($m_{ck}$) is very similar to $m_{c,MBASS}$ +3$\sigma_{MBASS}$ (Fig. 9), which suggests that the ALMM provides conservative estimates of the catalogue completeness threshold. In comparison, the link between the Ogata-Katsura $m_c$ proxy and max($m_{ck}$) depends on the catalogue, which could be explained by the curvature of the Ogata-Katsura model not always reflecting the observed FMD shape. The ALMM being more consistent with MBASS than with the Ogata-Katsura model suggests that the complete part of the mixture model is less biased than the Ogata-Katsura model by the FMD shape below completeness. The Ogata-Katsura model is however less sensitive than the ALMM to rounding errors in catalogues. The challenge of estimating $m_c$ for a class IV FMD, or 'broad-peaked' FMD, was demonstrated by Roberts et al. (2015), leading the authors to propose a 'best practice' workflow that combined different FMD-based $m_c$ estimators with $m_c$ and $b$-value error threshold rules. We would however follow the recommendation of Mignan & Chouliaras (2014), which is to estimate the catalogue completeness from max($m_c$) obtained from spatial mapping (e.g., BMC mapping) or from $m_c$+3$\sigma$ obtained from an estimator that is unsensitive to lack or not of curvature in the bulk FMD (MBASS is one example; Mignan et al., 2011; Mignan & Chouliaras, 2014). Despite the apparent flexibility of the ALMM to fit a variety of FMD shapes (compare Figs. 7-8 to 9), it remains subject to unexpected problems in the incomplete part of the catalogue. Simpler methods should therefore be preferred if the goal of the study is only to estimate $m_c$ and the $b$-value (and not make use of the incomplete data).

We finally observe that the mean $k$-value is relatively stable at c. 3, in agreement with the preliminary results of Mignan (2012). Exceptions include France, which could be explained by the rounding problems, and the ISC-GEM global catalogue, where an artificial cut-off seems to have been applied. In the case of a hard cut-off, *k* would indeed tend towards infinity. Mignan & Chen (2016) suggested a link between *k* and the seismic noise amplitude distribution, but this has yet to be demonstrated.

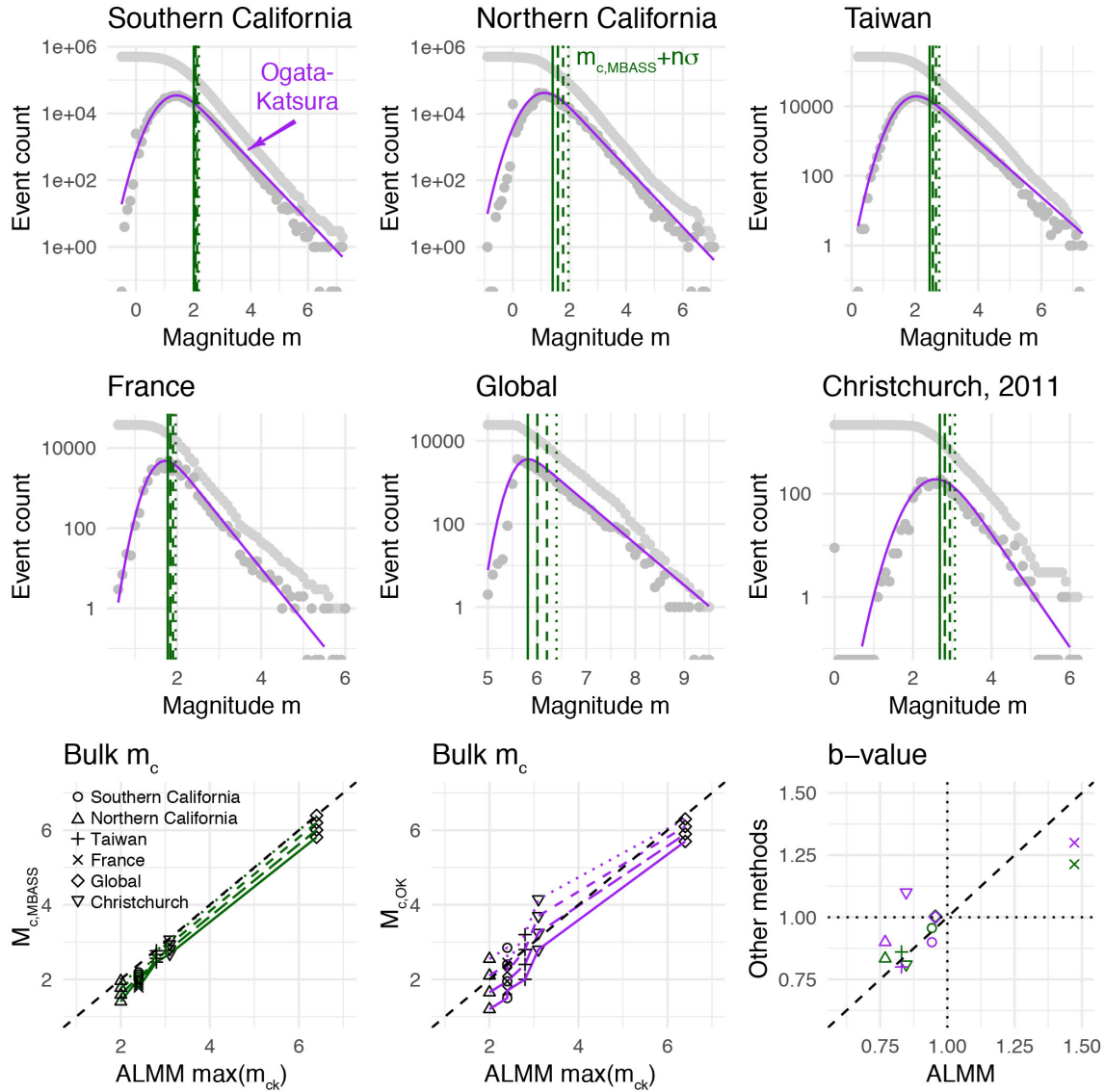

**Figure 9.** Comparison of the ALMM parameter estimates to results of other methods. The values $n = \{0, 1, 2, 3\}$ are represented by solid, long-dashed, dashed, and dotted lines, respectively. Note that the curvature of the Ogata-Katsura model does not always reflect the observed FMD shape in contrast with the ALMM, which is more flexible but alas more subject to rounding errors (see Figs. 7-8).

## 5 Conclusions

We presented an asymmetric Laplace mixture model of the generalized frequency magnitude distribution of earthquakes (so-called GFMD-ALMM). Despite some inherent limitations to correctly estimate the number $K$ of eFMDs for low-weight components and some sensitivity of the model to rounding errors in the incomplete part of earthquake catalogues, the main parameters (detection parameter κ, Gutenberg-Richter β-value, and $m_c$ range) have been shown to be reasonably retrieved for different FMD classes (Figs. 3-4, 6), in both simulations and real catalogues (Figs. 6-8). These results suggest that there is no need to discard events below $m_c$ since a mixture model can fit the various shapes that incomplete seismicity data may take. The proposed FMD mixture model should be seen as a complementary approach to template matching and network densification (e.g. Gibbons & Ringdal, 2006; Kraft et al., 2013), but with the advantage of being directly applicable to conventional earthquake catalogues, which are widely available.

The GFMD-ALMM has thus the potential, at least in theory, to improve precursory seismicity studies (e.g. Mignan, 2014). Indeed, up to 90% of seismicity data (ratio of events with $m < \max(m_c)$) is potentially discarded in regional seismicity analyses (c. 91% in class IV simulations, 93% and 90% calculated for the southern California and Nevada datasets of section 4.1, and between 83% and 90% for the regional catalogues of section 4.2). One could also imagine applying the ALMM to template matching catalogues to further increase the number of events available for statistical analysis. However, one would still need to make sure that the data set remains homogeneous when including incomplete data. Habermann (1982) pioneered this concept from the point of view of earthquake prediction studies, stating that: "*a data set in which a constant portion of the events in any magnitude bend are consistently reported through time is crucial for the recognition of seismicity rate changes which are real (related to some process change in the earth). Such a data set is termed a homogeneous data set.*" The GFMD-ALMM adds the [min($m$), $m_c$) "*magnitude band*" to the standard [$m_c$, max($m$)] band. An estimation of its parameter set $\theta_k$ in different time windows would allow verifying if the incomplete data is homogeneous or not in any specific earthquake predictability study. It remains unclear whether the possible parameter biases due to rounding errors in the incomplete part of the data, as observed in northern California and France, would pose problem to determine the 'homogeneity' of the catalogues over time.

Future improvements of the GFMD-ALMM could include an increased semi-supervision of the EM algorithm, by constraining the number of components $K$ from the seismic network spatial configuration (Eq. 13; Fig. 5), as well as the $m_c$ distribution shape from the BMC prior (Eq. 4). One could also investigate the impact of selecting the best bootstrap fits to potentially reduce parameter uncertainties. The practical advantages, if any, of the proposed mixture model in earthquake predictability research will be investigated elsewhere.

**Table 1.** FMD parameter estimates obtained by different methods.

| Catalogue | Parameter | ALMM | MBASS+3σ | Ogata-Katsura+3σ |
|---|---|---|---|---|
| Southern California | $\max(m_{ck})$ | 2.4 | 2.2 | 2.8 |
|  | $b$ | 0.94±0.04 | 0.96 | 0.9 |
|  | $k$ | 3.08±0.23 | N/A | N/A |
| Northern California | $\max(m_{ck})$ | 2.0 | 2.0 | 2.5 |
|  | $b$ | 0.77±0.05* | 0.83 | 0.9 |
|  | $k$ | 3.07±0.68* | N/A | N/A |
| Taiwan | $\max(m_{ck})$ | 2.8 | 2.8 | 3.2 |
|  | $b$ | 0.83±0.04 | 0.86 | 0.8 |
|  | $k$ | 3.20±0.28 | N/A | N/A |
| France | $\max(m_{ck})$ | 2.4 | 2.0 | 2.4 |
|  | $b$ | 1.47±0.14* | 1.21 | 1.3 |
|  | $k$ | 5.12±0.77* | N/A | N/A |
| Global | $\max(m_{ck})$ | 6.4 | 6.4 | 6.3 |
|  | $b$ | 0.95±0.07 | 1.00 | 1.0 |
|  | $k$ | 7.47±0.72 | N/A | N/A |
| Christchurch aftershocks | $\max(m_{ck})$ | 3.1 | 3.1 | 4.1 |
|  | $b$ | 0.84±0.06 | 0.81 | 1.1 |
|  | $k$ | 2.43±0.27 | N/A | N/A |

* Biased estimates likely due to rounding errors present in the incomplete part of the catalogues.